\documentclass[aps,prb,reprint,longbibliography,superscriptaddress]{revtex4-2}
\usepackage{mathrsfs}
\usepackage{amsmath,gensymb}
\usepackage{amsfonts}
\usepackage{amssymb}
\usepackage{amsthm}
\usepackage{graphicx}
\usepackage{natbib}
\usepackage{color}
\usepackage{hyperref}
\usepackage{bm}
\usepackage[caption=false]{subfig}
\usepackage{verbatim}
\usepackage[normalem]{ulem}

\begin{document}
\title{Oscillatory superconducting transition temperature in superconductor/antiferromagnet heterostructures}

\author{G. A. Bobkov}
\affiliation{Moscow Institute of Physics and Technology, Dolgoprudny, 141700 Moscow region, Russia}

\author{V. M. Gordeeva}
\affiliation{Moscow Institute of Physics and Technology, Dolgoprudny, 141700 Moscow region, Russia}

\author{A. M. Bobkov}
\affiliation{Moscow Institute of Physics and Technology, Dolgoprudny, 141700 Moscow region, Russia}

\author{I.V. Bobkova}
\affiliation{Moscow Institute of Physics and Technology, Dolgoprudny, 141700 Moscow region, Russia}
\affiliation{National Research University Higher School of Economics, 101000 Moscow, Russia}

\begin{abstract}
One of the most famous proximity effects at ferromagnet/superconductor (F/S) interfaces is partial conversion of singlet superconductivity to triplet pairing correlations. Due to the presence of macroscopic exchange field in the ferromagnet the Cooper pairs penetrating into the ferromagnet from the superconductor acquire a finite momentum there. The finite-momentum pairing manifests itself, in particular, as a nonmonotonic dependence of the critical temperature of the bilayer on the thickness of the F layer. Here we predict that despite the absence of the macroscopic exchange field the critical temperature of the antiferromagnet/superconductor (AF/S) bilayers also exhibit nonmonotonic (oscillating) dependence on the AF layer thickness. It is a manifestation of the proximity-induced N\'eel-type triplet correlations, which acquire finite total pair momentum and oscillate in the AF layer due to the Umklapp electron scattering processes at the AF/S interface. Our prediction can provide a possible explanation for a number of recently published experimental observations of the critical temperature of AF/S bilayers.
\end{abstract}

\maketitle

\section{Introduction}

Materials with magnetic ordering and superconductors both have prominent roles in condensed matter physics due to their fundamental and technological interest. In particular, magnets are key elements of various spintronic applications and superconductors manifest perfect diamagnetism and dissipationless transport. Such physics is interesting in itself and is implemented in these materials separately. However, when we combine these materials, new physics related to the nanoscale interface region between them can occur. In general it is called proximity effects. 

One of the most famous proximity effects at ferromagnet/superconductor (F/S) interfaces is partial conversion of singlet superconductivity to triplet pairing correlations \cite{Buzdin2005,Bergeret2005}. The triplet pairs can sustain dissipationless spin currents and, consequently, are cornerstone elements in superconducting spintronics \cite{Eschrig2015,Linder2015}. The triplet pairs arise at the expense of singlet correlations and thus suppress singlet superconductivity.     One of the important properties of the triplet pairs generated at F/S interfaces is the finite momentum of the pair \cite{Buzdin1982,Demler1997}, what allows them to be called a mesoscopic analogue of the inhomogeneous Fulde-Ferrel-Larkin-Ovchinnikov (FFLO) superconducting state \cite{Larkin1964,Fulde1964}. The finite momentum, which the Cooper pair acquires in
the exchange field of the ferromagnet, makes the pairing
wave function oscillating. The resulting phase change across
the ferromagnetic layer is responsible for the $\pi$-junction
effects \cite{Buzdin1982,Buzdin2005,Kontos2002,Ryazanov2001,Oboznov2006, Bannykh2009, Robinson2006}, which are widely used now in the superconducting electronics \cite{Yamashita2005,Feofanov2010,Shcherbakova2015}. The interference of the incident and reflected oscillating wave functions determines the oscillatory phenomena of the critical temperature $T_c$ versus the F layer thickness in F/S bilayers and multilayers, which have been widely studied both theoretically \cite{Fominov2002,Radovic1991,Vodopyanov2003,Lazar2000,Buzdin2000,Zareyan2001} and experimentally \cite{Jiang1995,Mercaldo1996,Muhge1996,Zdravkov2006,Zdravkov2010}.

At present, there is another direction of superconducting spintronics, which is based on superconductor/antiferromagnet (AF/S) heterostructures. It is often called antiferromagnetic superconducting spintronics and looks very promising due to several advantages brought by antiferromagnets, such as negligible stray fields and intrinsic high-frequency dynamics \cite{Baltz2018,Jungwirth2016,Brataas2020}, and also due to unique features of the proximity effect at AF/S interfaces \cite{Andersen2006,Enoksen2013,Bobkova2005,Andersen2005,Jonsen2021,Bobkov2022,Bobkov2023,Chourasia2023,Rabinovich2019,Falch2022,Jakobsen2020,Lado2018,Fyhn2022_1}, in particular, the possibility to control the anisotropy of the N\'eel vector by superconductivity \cite{Bobkov2023_2} and AF/S/AF spin valves \cite{Johnsen_Kamra_2023}. 

Analogously to F/S interfaces the singlet-triplet conversion was also reported even at fully compensated AF/S interfaces \cite{Bobkov2022}, but the amplitude of the corresponding oscillations flips its sign between the nearest neighbor sites of the lattice in the superconductor in the same way as the N\'eel magnetic order does. For this reason the corresponding triplet correlations were called N\'eel triplets. The influence of the N\'eel triplets on superconducting critical temperature has already been investigated in thin-film AF/S heterostructures \cite{Bobkov2022,Bobkov2023}, which can be viewed as homogeneous superconductors in uniform N\'eel exchange field. However, their behavior in metallic antiferromagnets of finite width, proximitized by a superconductor, has not been studied yet. In this work we fill this gap. Naively, one does not expect that a Cooper pair penetrating into the antiferromagnet from the superconductor possesses a finite total momentum because the average value of the exchange field in the antiferromagnet is zero, the quasiparticles spectrum is spin-degenerate and, therefore, spin-up and spin-down electrons, forming the pair, should have opposite momenta with equal absolute values $\bm p_\uparrow = -\bm p_\downarrow$. In its turn, that means zero total momentum of the pair and, as a result, absence of the oscillations of the pair amplitude. Consequently, one does not expect oscillating behavior of the critical temperature of AF/S bilayers in dependence on the thickness of the antiferromagnetic layer. Indeed, in the regime, when the N\'eel triplets can be disregarded, this dependence has been calculated and no oscillations were reported \cite{Fyhn2022_1}.

However, there are a number of experimental works, where the critical temperature of AF/S bilayers with metallic antiferromagnets has been measured as a function of the AF thickness and the oscillating behavior was observed \cite{Bell2003,Hubener2002,Wu2013}. Here we demonstrate theoretically that taking into account the N\'eel triplet correlations at AF/S interfaces results in the oscillating dependence of the critical temperature on the AF thickness and unveil physical mechanisms of the effect. Thus, oscillations of the critical temperature of AF/S bilayers can be viewed as a signature of the presence of N\'eel-type triplet correlations in the heterostructure.

The paper is organized as follows. In Sec.~\ref{model} we describe the considered model and the formalism used for calculations. In Sec.~\ref{results} our results are presented: Sec.~\ref{oscillations_f} is devoted to discussion of spatial oscillations of the triplet pair correlations, induced in the AF layer by proximity to the superconductor, and in Sec.~\ref{oscillations_Tc} we show the results for the critical temperature and discuss how it depends on the parameters of the AF/S heterostructure. Our conclusions are presented in Sec.~\ref{conclusions}. In the Appendix we provide some technical details of the calculation of the Green's function.

\section{Model and method}
\label{model}

\begin{figure}[tb]
	\begin{center}
		\includegraphics[width=65mm]{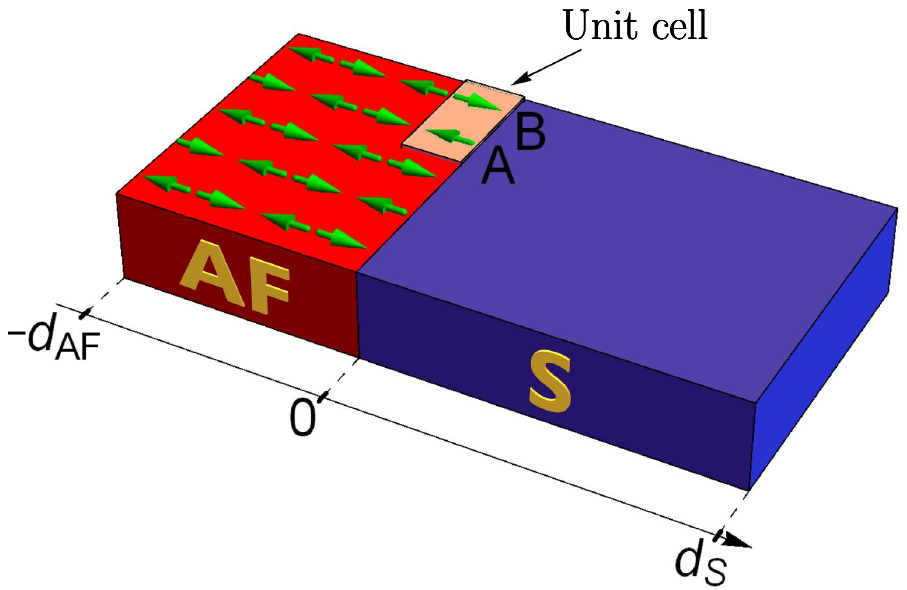}
		\caption{AF/S bilayer. Staggered magnetization of the AF layer is schematically depicted by arrows. The unit cell containing two sites belonging to A and B sublattices is also shown.}
  \label{fig:sketch}
	\end{center}
 \end{figure}

We consider an AF/S bilayer system, presented in Fig.~\ref{fig:sketch}. It is composed of a conventional $s$-wave singlet superconductor with thickness $d_S$ and a metallic antiferromagnet with thickness $d_{AF}$. The system is described by the following tight-binding Hamiltonian in the two-sublattice representation:
\begin{align}
\hat H= - &t \sum \limits_{\langle \bm{i}\bm{j}\nu \bar \nu\rangle ,\sigma} \hat \psi_{\bm{i} \sigma}^{\nu\dagger} \hat \psi_{\bm{j} \sigma}^{\bar \nu} + \sum \limits_{\bm{i},\nu } (\Delta_{\bm{i}}^\nu \hat \psi_{\bm{i}\uparrow}^{\nu\dagger} \hat \psi_{\bm{i}\downarrow}^{\nu\dagger} + H.c.) - \nonumber \\
&\mu \sum \limits_{\bm{i} \nu, \sigma} \hat n_{\bm{i}\sigma}^\nu  
+ \sum \limits_{\bm{i} \nu,\alpha \beta} \hat \psi_{\bm{i}\alpha}^{\nu \dagger} (\bm{h}_{\bm{i}}^\nu \bm{\sigma})_{\alpha \beta} \hat \psi_{\bm{i}\beta}^\nu.
\label{ham_2}
\end{align}
The unit cell with two sites - A and B - is introduced as shown in Fig.~\ref{fig:sketch}. In the framework of this two-sublattice approach the unit cells as a whole are marked by radius-vector $\bm i$. $\nu=A,B$ is the sublattice index, $\bar \nu = A(B)$ if $\nu=B(A)$ means that the corresponding quantity belongs to the opposite sublattice. $\langle \bm i \bm j \nu \bar \nu\rangle $ means summation over the nearest neighbors, $\hat{\psi}_{\bm i \sigma}^{\nu \dagger}(\hat{\psi}_{\bm i \sigma}^{\nu })$ is the creation (annihilation) operator for an electron with spin $\sigma$ at the sublattice $\nu$ of the unit cell $\bm i$. $t$ parameterizes the hopping between adjacent sites, $\mu$ is the electron chemical potential counted from the middle of the conduction band (that is $\mu=0$ corresponds to half-filling), and $\hat n_{\bm i \sigma}^\nu = \hat \psi_{\bm i \sigma}^{\nu\dagger} \hat \psi_{\bm i \sigma}^\nu$ is the particle number operator at the site belonging to sublattice $\nu$ in unit cell $\bm i$. For simplicity the hopping parameter is assumed to be equal in the regions occupied by the supersonductor and by the antiferromagnet. At the same time, we take into account difference between chemical potentials in the S and AF regions $\mu_S \neq \mu_{AF}$. It is worth mentioning that for the equilibrium problem under consideration the filling levels of electronic states in the S and AF layers are the same. Different $\mu_{S}$ and $\mu_{AF}$ mean that conduction bands of the materials are shifted relative to each other. The staggered magnetism is described by $\bm h_{\bm i}^{A(B)} = + (-) \bm h_{\bm i}$, where $\bm h_{\bm i}$ is the local magnetic moment at site A of the unit cell with the radius-vector $\bm i$ in the AF.  This allows us to consider $\bm h_{\bm i}$  as a slow function of the spatial coordinate. We consider the homogeneously ordered N\'eel state of the AF here, such that $\bm h_{\bm i} = \bm h$ does not depend on the position $\bm i$ inside the antiferromagnet and has zero value in the superconductor. We assume that the AF/S interface is fully compensated, that is the interface exchange field is staggered with zero average value. $\Delta_{\bm{i}}^\nu$ accounts for on-site $s$-wave pairing and has nonzero value only in the superconductor. In case of conventional singlet pairing $\Delta_{\bm{i}}^A = \Delta_{\bm{i}}^B = \Delta_{\bm{i}}$.

The system, described by Hamiltonian (\ref{ham_2}), can be treated in the framework of the two-sublattice quasiclassical theory \cite{Bobkov2022}. In this formalism all the characteristic spatial scales are large as compared to interatomic distance and, therefore, the discrete unit cell index $\bm i$ can be changed by the continuous spatial variable $\bm R$. The system is described by the quasiclassical Green's function $\check g(\bm R, \bm p_F, \omega_m)$, which is a $8 \times 8$ matrix in the direct product of spin, particle-hole and sublattice spaces and depends on the radius-vector $\bm R$, direction of the electron momentum at the Fermi surface $\bm p_F$ and fermionic Matsubara frequency $\omega_m = \pi T(2m+1)$. It obeys the following  Eilenberger equation \cite{Bobkov2022}:
\begin{align}
&\Bigl[ \Bigl(i \omega_m \tau_z + \mu + \tau_z \check \Delta(\bm R) -\bm h(\bm R) \bm\sigma\tau_z \rho_z\Bigr)\rho_x, \check g(\bm R, \bm p_F,\omega_m) \Bigr]  \nonumber \\
&+i \bm v_F \bm \nabla \check g(\bm R, \bm p_F,\omega_m) = 0,
\label{Eilenberger}
\end{align}
where $\bm v_F$ is the Fermi velocity for the trajectory $\bm p_F$, $\bm \sigma = (\sigma_x, \sigma_y, \sigma_z)^T$ is the vector of Pauli matrices in spin space. Analogously, $\tau_i$ and $\rho_i$ are Pauli matrices in particle-hole and sublattice spaces, respectively. As it is shown in Fig.~\ref{fig:sketch}, the AF layer occupies region $-d_{AF}<x<0$, and the S layer is at $0<x<d_S$. We assume translational invariance along the AF/S interface, consequently the Green's function only depends on $x$-coordinate, normal to the interface. For brevity we define $|\bm v_{F,x}|=v$. $\check \Delta(\bm R) = \Delta(\bm R) \tau_x$. The quasiclassical Green's function for a given trajectory $\check g(x,\pm v) \equiv \check g^{\pm}(x)$ also obeys the normalization condition:
\begin{align}
     \left[\check g^{\pm}(x)\right]^2=1
\label{eq:normalization_condition}
\end{align}
and boundary conditions at the AF/S interface $x=0$ and at the impenetrable edges of the AF ($x=-d_{AF}$) and S ($x=d_S$) layers.
Since we consider identical lattices in the AF and S layers and $(\mu_{AF}, \mu_S)$ are assumed to be small with respect to $t$ to be considered in the framework of the quasiclassical approximation, the interface barrier at $x=0$ is absent. In this case the boundary conditions at $x=0$ are reduced to continuity of the Green's functions, analogously to the case of superconductor/normal metal and superconductor/ferromagnet interfaces \cite{Zaitsev1984,mrs1988}:
\begin{align}
    \check g^{S,\pm}(x=0)=\check g^{AF,\pm}(x=0).
    \label{eq:bs_interface}
\end{align}
The boundary conditions at the impenetrable edges are reduced to the equality of the incident and reflected Green's functions \cite{Zaitsev1984}:
\begin{align}
\begin{cases}
    \check g^{S,+}(x=d_S)=\check g^{S,-}(x=d_S) \\
    \check g^{AF,+}(x=-d_{AF})=\check g^{AF,-}(x=-d_{AF})
\end{cases}
\end{align}
 We assume $d_S$ to be much smaller than all the characteristic spatial scales of the anomalous Green's function in the superconductor. One of them is the superconducting coherence length $\xi_S = v_F/2\pi T_{c0}$ and the other one is the length of spatial oscillations of the triplet components of the anomalous Green's function, which will be discussed below. In this case the superconducting order parameter is approximately spatially constant inside the S layer, $\Delta (\bm R) = \Delta$. 

For the problem under consideration the N\'eel vector of the AF is spatially homogeneous. In this case the Green's function is diagonal in spin space and further we work with its components in spin space $\check g_\sigma$, which are still $4 \times 4$ matrices in the direct product of particle-hole and sublattice spaces. $\sigma = \uparrow, \downarrow$ in the subscripts of all the quantities and $\sigma = \pm 1$ as a factor in the mathematical expressions for spin-up (down) components, respectively. To find the critical temperature of the system we linearize the Eilenberger equation (\ref{Eilenberger}) with respect to $\Delta/T_c$. In this approximation the Green's function takes the form:
\begin{align}
\check g_\sigma = 
\left(
\begin{array}{cc}
\check g_{\sigma,N} & \check f_\sigma \\
\check {\tilde f}_\sigma & \check {\tilde g}_{\sigma,N}
\end{array}
\right)_{\tau}.
\label{gf_linearized}    
\end{align}
The diagonal components $\check g_{\sigma,N}$ and $\check {\tilde g}_{\sigma,N}$  are to be calculated in the normal state of the superconductor. The anomalous components $\check f_\sigma$ and $\check {\tilde f}_\sigma$ are of the first order with respect to $\Delta/T_c$. The detailed calculations of the normal state Green's function $\check g_{\sigma,N}$ are presented in Appendix. For the problem under consideration in the AF layer it takes the form:
\begin{align}
\check g_{\sigma,N}^{AF} &= A\left[\rho_x - \frac{i \sigma h}{\mu_{AF}+i\omega_m}\rho_y\right] \nonumber \\
&+  B \left[\frac{i \sigma h}{\Lambda} \rho_x + \frac{\mu_{AF}+i \omega_m}{\Lambda}\rho_y + \rho_z \right]e^{\beta x} \nonumber \\
&+  C \left[-\frac{i \sigma h}{\Lambda} \rho_x - \frac{\mu_{AF} +i \omega_m}{\Lambda}\rho_y + \rho_z \right]e^{-\beta x},
\label{eq:g_N_AF}
\end{align}
where $h$ is the absolute value of $\bm h$, $\Lambda=\sqrt{h^2-(\mu_{AF}+i\omega_m)^2}$, $\beta=2\Lambda/v$, and $A,B,C$ - coefficients, which are found from the boundary and normalization conditions. In the S layer the Green's function takes the form:
\begin{align}
\check g_{\sigma,N}^{S} &= G \rho_x +  D \left[\rho_z + i \rho_y \right]e^{\lambda x} +  E \left[\rho_z - i \rho_y \right]e^{-\lambda x},
\label{eq:g_N_S}
\end{align}
where $\lambda=2(\omega_m-i\mu_S)/v$, and $G, D,E$ - coefficients, which are found from the boundary and normalization conditions.

The hole component $\check {\tilde g}_N$ is obtained as follows:  
\begin{align}
    \check {\tilde g}_{\sigma, N}(\omega_m,h,\mu)=-\check g_{\sigma, N} (-\omega_m,-h,\mu).
    \label{eq:hole_normal}
\end{align}
It is worth noting that unlike F/S structures for AF/S systems the normal state Green's function $\check g_N$ is not identically equal to 1 and has complicated structure with spatially constant and spatially oscillating components with strongly energy-dependent amplitudes. This originates from the energy dependence of the density of states near the Fermi surface because of the antiferromagnetic gap and possibility of Umklapp processes, see qualitative discussion of the nature of the oscillations below.

The anomalous components $\check f_\sigma$ and $\check {\tilde f}_\sigma$ are of the first order with respect to $\Delta/T_c$.  The resulting equation for the anomalous Green's function $\check f_\sigma$ takes the form:
\begin{align}
\left\{ (i \omega_m  -  h(x) \sigma \rho_z )\rho_x, \check f^{\pm}_\sigma \right\} + \left[ \mu(x) \rho_x, \check f^{\pm}_\sigma \right]+ \\ \nonumber
+\Delta(x) \left( \rho_x \check {\tilde g}_N^{\pm} - \check g_N^{\pm} \rho_x \right)\pm i v \frac{d}{dx} \check f^{\pm}_\sigma= 0.
\label{f_linearized_eq}    
\end{align}
This equation can be rewritten in terms of the vector $\hat f^{\pm}_\sigma=(\check f_{\sigma,0}^{\pm},\check f_{\sigma,x}^{\pm},\check f_{\sigma,y}^{\pm},\check f_{\sigma,z}^{\pm})^T$, where the components of the vector correspond to the components of the expansion of $\check f_\sigma^{\pm} = \sum \limits_i \check f_{\sigma,i}^\pm \rho_i$ over the Pauli matrices in the sublattice space: 
\begin{eqnarray}
    \hat F_\sigma \hat f^{\pm}_\sigma +G^{\pm}_\sigma \pm iv \frac{d}{dx} \hat f^{\pm}_\sigma=0,
    \label{linearized_vector}
\end{eqnarray}
\begin{align}
    \hat F_\sigma= \left(
\begin{array}{cccc}
0 & 2i\omega_m & -2i h(x) \sigma & 0 \\
2i\omega_m & 0 & 0 & 0 \\
-2i h(x) \sigma & 0 & 0 & -2i\mu(x) \\
0 & 0 & +2i\mu(x) & 0 
\end{array}
\right).
\end{align}
The same expansion $\check g_{\sigma,N}^{\pm} = \sum \limits_i \check g_{\sigma,N,i}^\pm \rho_i$ is also introduced for the vector of the normal state Green's function. Then
\begin{align}
\hat G^{\pm}_\sigma=\left(
\begin{array}{c}
\Delta(x)(\check{\tilde g}_{\sigma,N,x}^{\pm}-\check g_{\sigma,N,x}^{\pm})  \\
0 \\
-i\Delta(x)(\check{\tilde g}_{\sigma,N,z}^{\pm}+\check g_{\sigma,N,z}^{\pm}) \\
i\Delta(x)(\check{\tilde g}_{\sigma,N,y}^{\pm}+\check g_{\sigma,N,y}^{\pm})
\end{array}
\right).
\label{linearized_vector_N}
\end{align}
We assume that the S-layer is thin  with $d_S\ll\xi_S$, then we can linearize Eq.~(\ref{linearized_vector}) with respect to $x/\xi_S$ and at the same time it is possible to consider the superconducting order parameter in the S layer $\Delta(x)$  independent of coordinates $\Delta(x) \approx \Delta$. With this assumption,  the solution of Eq.~(\ref{linearized_vector}) takes the following form in the AF and S regions, respectively (the boundary conditions at $x=+d_S,-d_{AF}$ are already taken into account):
\begin{align}
    \hat f_\sigma^{\pm,AF}=\sum\limits_{\alpha=1}^4 K_{\alpha,\sigma}^\pm\hat e_{\alpha} e^{\mp i \kappa_{\alpha}(x+d_{AF})/v},
    \label{eq:anomalous_F}
\end{align}

\begin{align}
    \hat f_\sigma^{\pm,S}=\sum\limits_{\alpha=1}^4 L_{\alpha,\sigma}^\pm\hat s_{\alpha} e^{\mp i \nu_{\alpha}(x-d_{S})/v}-\frac{(\hat G_\sigma^\pm \hat s_{\alpha})\hat s_{\alpha}}{\nu_{\alpha}|s_{\alpha}|^2},
    \label{eq:anomalous_S}
\end{align}
where $\hat e_{\alpha},\hat s_{\alpha}$ are eigen vectors of matrix $\hat F_\sigma^\pm$ in the AF and S regions, $\kappa_{\alpha},\nu_{\alpha}$ are eigen values of matrix $\hat F_\sigma^\pm$ in the AF and S regions, respectively. $K_{\alpha,\sigma}^\pm,L_{\alpha,\sigma}^\pm$ are unknown coefficients, which should be found from the boundary conditions at $x=0$. 
\begin{align}
    \hat e_{1,2}=\left(
\begin{array}{c}
\mu_{AF}  \\
\pm \sqrt{\mu_{AF}^2-h^2} \\
0 \\
-h \sigma
\end{array}
\right),~\hat e_{3,4}=\left(
\begin{array}{c}
-h \sigma \\
0  \\
\mp i\sqrt{\mu_{AF}^2-h^2} \\
+\mu_{AF}
\end{array}
\right),
\end{align}

\begin{align}
    \hat s_{1,2}=\left(
\begin{array}{c}
1  \\
\pm 1 \\
0 \\
0
\end{array}
\right),~\hat s_{3,4}=\left(
\begin{array}{c}
0  \\
0 \\
\mp i \\
1
\end{array}
\right),
\end{align}

\begin{align}
    \kappa_{1,2}=\pm 2i\omega_m \frac{\mu_{AF}}{\sqrt{\mu_{AF}^2-h^2}},~\kappa_{3,4}=\pm 2\sqrt{\mu_{AF}^2-h^2},
\end{align}

\begin{align}
    \nu_{1,2}=\pm 2i\omega_m,~\nu_{3,4}=\pm 2 \mu_S.
\end{align}
The exact expressions for $L_{\alpha,\sigma}$ and $K_{\alpha,\sigma}$ are rather lengthy and we do not write them here. However, for the case $\mu_S = 0$, for which we present numerical results, the coefficients are written in  Appendix.

The critical temperature of the AF/S bilayer is calculated from the self-consistency equation
\begin{align}
\Delta(x) = \int \frac{d\Omega}{4\pi} i \pi \lambda  T_c \sum \limits_{\omega_m} f_s(x) ,
\label{Tc}    
\end{align}
where $\int \frac{d\Omega}{4\pi}$ means averaging over the Fermi surface, $\lambda$ is coupling constant, $f_s$ - amplitude of singlet correlations, which takes the form:
\begin{align}
    f_s^\pm=\sum \limits_\sigma {\rm Tr}[\frac{\rho_x \check f_\sigma^\pm}{4}]=\sum \limits_\sigma\left [ (L_{1,\sigma}^\pm-L_{2,\sigma}^\pm)-\frac{\hat G_\sigma^\pm(\hat s_1+\hat s_2)}{4i\omega_m} \right ].
    \label{eq:f_singlet_result}
\end{align}
At $d_S \lesssim \xi_S$ the spatial dependence of the order parameter in the S layer is weak, that is $\Delta(x) \approx \Delta$. In this case $f_s(x)$ in Eq.~(\ref{Tc}) can be taken at $x=d_S$ or we can take the average of the $f_s$ over the S layer. The result depends little on this. 

\section{Results}
\label{results}

\subsection{Oscillations of the N\'eel triplet correlations in the antiferromagnet}
\label{oscillations_f}

\begin{figure}[tb]
	\begin{center}
		\includegraphics[width=85mm]{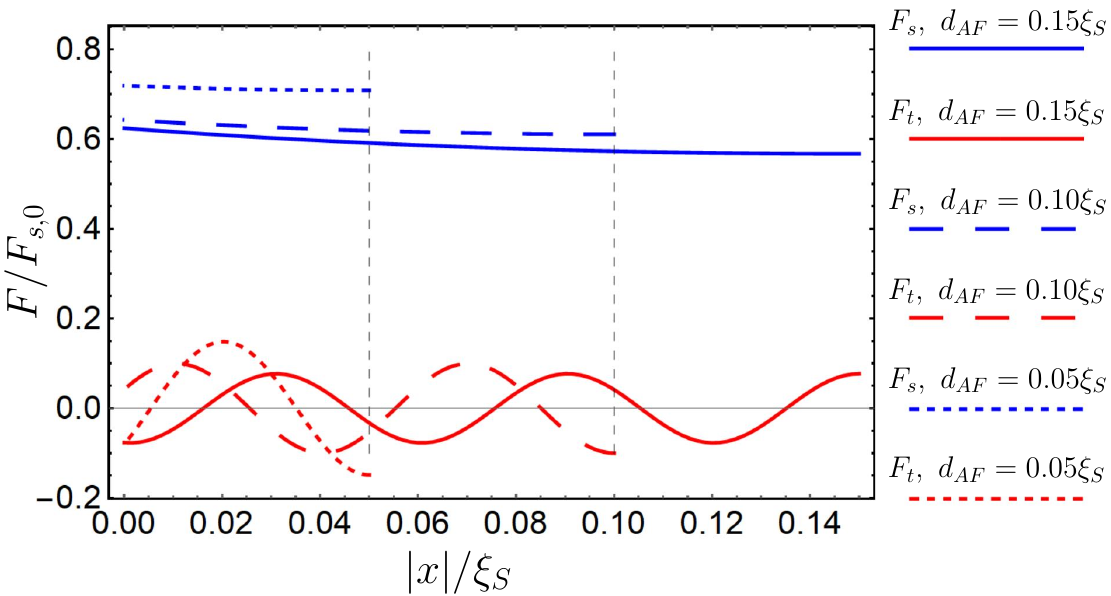}
\caption{Dependence of the triplet anomalous Green's function $F_t$(red) and the singlet anomalous Green's function $F_s$(blue) for the normal to the AF/S interface trajectory $v>0$ on the distance from the  interface inside the AF layer. Different curves correspond to different thicknesses $d_{AF}$ of the AF layer. Each of the curves ends at the distance corresponding to the impenetrable edge of the AF layer.  $d_S=0.5\xi_S$, $h=0.6\mu_{AF}$, $\mu_S=0$, $\mu_{AF}=420 T_{c0}$, where $T_{c0}$ is the value of the critical temperature without the AF layer. $F_{s,0}$ is the singlet anomalous Green's function in the absence of the AF layer, see text for the details of the definitions of the Green's functions.}
 \label{fig:oscillations_anomalous}
	\end{center}
\end{figure}

Now we discuss the behavior of the singlet and N\'eel triplet correlations in the AF layer. Fig.~\ref{fig:oscillations_anomalous} shows some typical examples of the spatial distribution of the singlet and triplet correlations inside the AF layer. The anomalous Green's functions plotted in this figure are summed up over all positive Matsubara frequencies, that is
\begin{align}
    F_{s,t} = T \sum \limits_{\omega_m > 0}f_{s,t},
    \label{F_s_sum}
\end{align}
with
\begin{align}
    f_t = \frac{1}{4}\sum \limits_\sigma \sigma {\rm Tr}[\rho_y \check f_\sigma^\pm],
    \label{eq:f_triplet}
\end{align}
where $T \to T_c$. The anomalous Green's functions $F_{s,t}$ are normalized to the value of the singlet anomalous Green's function $F_{s,0}$ corresponding to the isolated S layer without proximity to an antiferromagnet. The correlations presented in Fig.~\ref{fig:oscillations_anomalous} correspond to the $v>0$ trajectory normal to the AF/S interface. 

\begin{figure}[tb]
	\begin{center}
		\includegraphics[width=75mm]{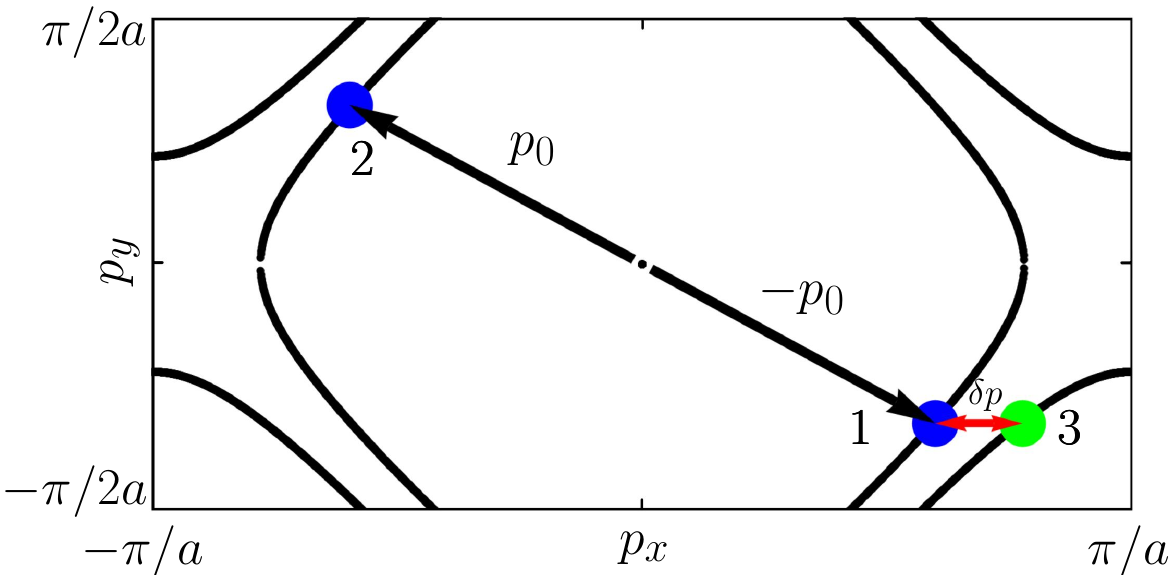}
		\caption{BZ and Fermi surface (black curves) of the AF layer. Zero-momentum Cooper pair between electrons 1 and 2 is schematically shown by black arrows. There is also N\'eel-type finite-momentum triplet pairing between electrons 2 and 3, which is produced from electron 1 due to the Umklapp reflection process from the AF/S interface, see text for details. The total momentum of the pair (2,3) $\delta p$ is shown by the red arrow.}
  \label{fig:explanation}
	\end{center}
\end{figure}

From the results presented in Fig.~\ref{fig:oscillations_anomalous} we conclude that the triplet correlations oscillate inside the antiferromagnet, while the singlet correlations just decay without oscillations. Now we discuss the physical explanation of the oscillating behavior. Let us consider an electron $(p_{x1}, p_y)$ incoming to the AF/S interface from the AF side (marked by 1 in Fig.~\ref{fig:explanation}). Due to proximity effect with the adjacent S layer it forms a singlet Cooper pair with another electron 2 corresponding to the momentum $(-p_{x1}, -p_y)$. For the plane interface the component of the electron momentum along the interface $p_y$ is conserved. For the interface problem under consideration it is convenient to choose the Brillouin zone as shown in Fig.~\ref{fig:explanation}. Due to the doubling of the unit cell the Brillouin zone (BZ) is compressed twice in the interface direction. As a result, additional branches of the Fermi surface appear in the reduced BZ and the incoming electron 1 can be reflected as electron 3, corresponding to the same $p_y$ (Umklapp process or the so-called Q-reflection \cite{Bobkova2005,Andersen2005}). Due to the proximity-induced pairing correlations between electrons 1 and 2, the pairing is also established between electrons 2 and 3. It is the N\'eel-type triplet finite-momentum pairing and the total momentum of the pair (2,3) can be found as $\delta p = |p_{x3}-p_{x1}|$. 

The normal state electron dispersion in the reduced BZ takes the form $\varepsilon = -\mu_{AF} + \sqrt{h^2 + 4t^2(\cos p_x a + \cos p_y a + \cos p_z a)^2}$. From this dispersion relation we obtain $\delta p = \sqrt{\mu_{AF}^2 - h^2}/(ta \sin [p_x a])$. The last expression can be rewritten in terms of the electron Fermi velocity $v_{F,x} \equiv v = \partial \varepsilon/\partial p_x = 2ta \sin [p_x a]$ at $\mu_{AF}=h=0$, which enters our quasiclassical theory, as $\delta p = 2\sqrt{\mu_{AF}^2 - h^2}/v$. Please note that it is exactly the factor which determines oscillation behavior of the AF anomalous Green's function in Eq.~(\ref{eq:anomalous_F}). Then the period of oscillations is
\begin{eqnarray}
L_{osc} = \frac{\pi v_F}{\sqrt{\mu_{AF}^2 - h^2}},
\label{eq:period}
\end{eqnarray} 
which is in agreement with the result of rigorous calculation of the anomalous Green's function, plotted in Fig.~\ref{fig:oscillations_anomalous}. 

Unlike the case of S/F heterostructures, where both singlet and triplet correlations manifest oscillations with the same period inside the ferromagnet, here the singlet correlations do not oscillate. The reason is that according to our qualitative consideration only N\'eel pairs can have finite momentum of such physical origin. However, in the considered case the N\'eel singlet pairs are not produced because of the absence of an appropriate scalar generator of the N\'eel type.

\subsection{Critical temperature of the AF/S bilayer}
\label{oscillations_Tc}

In AF/S bilayer systems with finite-width antiferromagnets the oscillating N\'eel triplet superconducting correlations discussed above can experience constructive or destructive interference due to the reflections from the impenetrable edge of the AF layer. It leads to the oscillating dependence of the N\'eel triplet correlations amplitude as a function of $d_{AF}$. In its turn, such a nonmonotonic dependence of the triplet amplitude results in the oscillating behavior of the critical temperature of the bilayer as a function of the AF layer width $d_{AF}$. 

\begin{figure}[tb]
	\begin{center}
		\includegraphics[width=85mm]{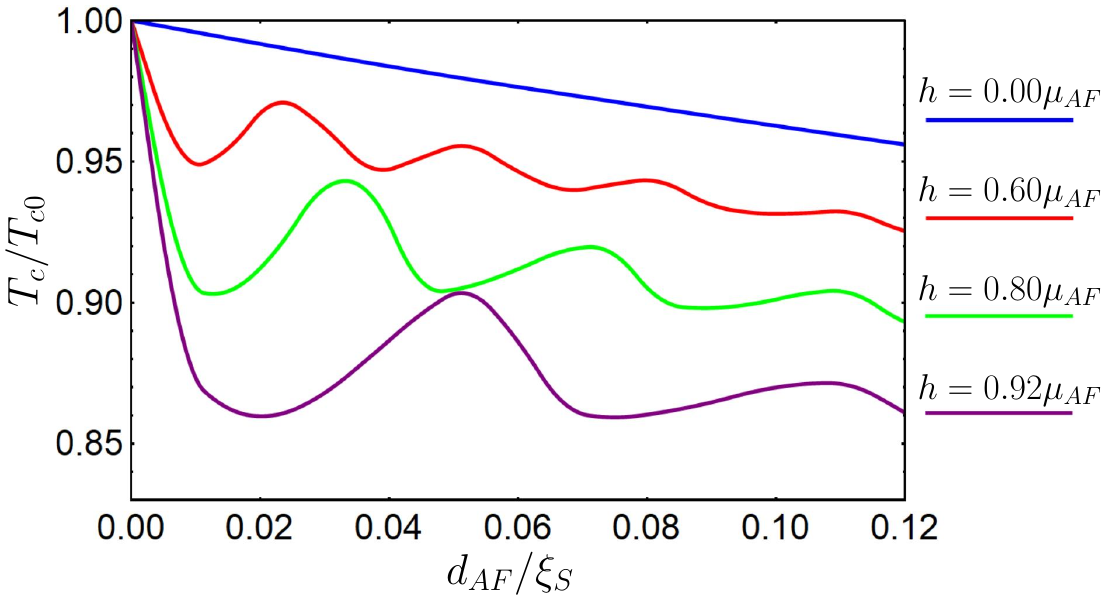}	
  \caption{Critical temperature of the AF/S bilayer as a function of the AF layer thickness $d_{AF}$. The critical temperature is normalized to its value without the AF layer $T_{c0}$. $d_S=1.5\xi_S$, $\mu_S=0$, $\mu_{AF}=420 T_{c0}$.}
  \label{fig:Tc1}
	\end{center}
\end{figure}

\begin{figure}[tb]
	\begin{center}
		\includegraphics[width=85mm]{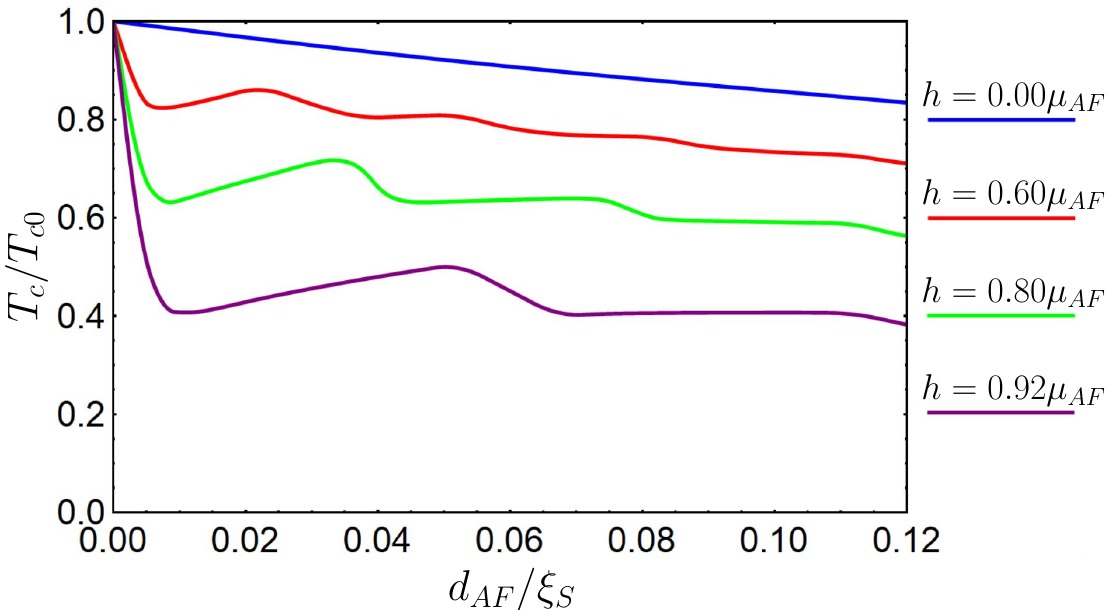}
		\caption{The same as in Fig.~\ref{fig:Tc1} but for $d_S=0.75\xi_S$. The other parameters are the same.}
  \label{fig:Tc2}
	\end{center}
\end{figure}

\begin{figure}[tb]
	\begin{center}
		\includegraphics[width=85mm]{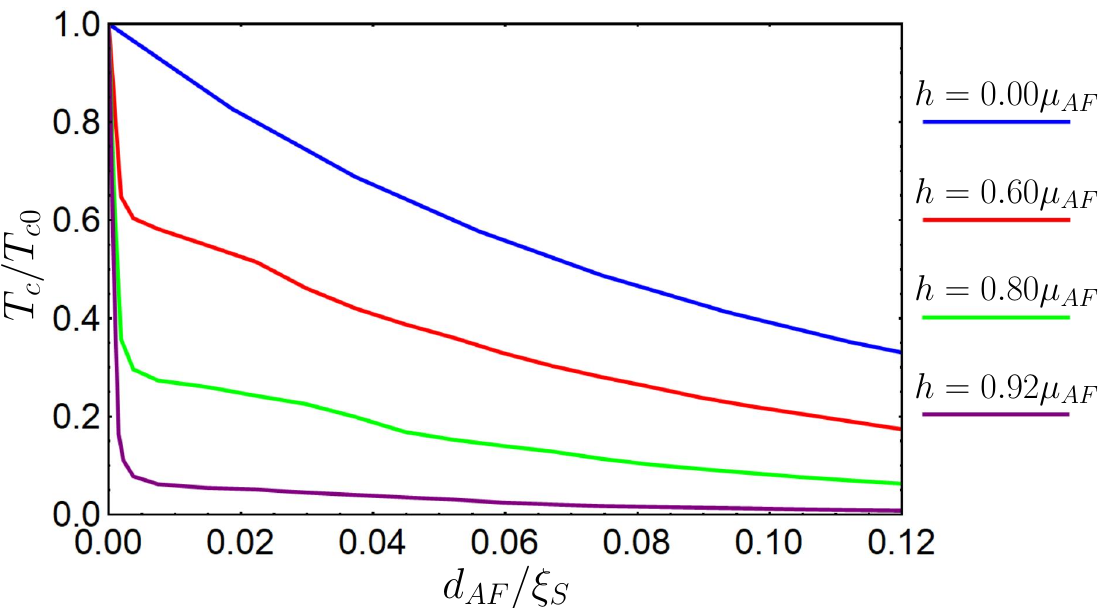}
		\caption{The same as in Fig.~\ref{fig:Tc1} but for $d_S=0.225\xi_S$. The other parameters are the same.}
  \label{fig:Tc3}
	\end{center}
\end{figure}

The results of the calculation of the critical temperature of the AF/S bilayer as a function of $d_{AF}$ are presented in Figs.~\ref{fig:Tc1}-\ref{fig:Tc3}. Different curves correspond to different exchange fields of the AF layer. In all three figures we can see the superconductivity suppression accompanied by  oscillations of the critical temperature. The amplitude of the oscillations grows with the value of the exchange field.  Fig.~\ref{fig:Tc1} demonstrates the results corresponding to rather thick superconductor layer. In this case the overall suppression of superconductivity is weak (pay attention to the scale of the vertical axis). It is obvious that in the limit of very thick S layer all the manifestations of the proximity effect between S and AF in the critical temperature (that is, the suppression and the oscillations) vanish. In Fig.~\ref{fig:Tc2} the results for the moderate thickness of the S layer are presented. The suppression of superconductivity as well as oscillations are more pronounced. In Fig.~\ref{fig:Tc3} the results for the thinnest S layer are shown. The suppression of superconductivity is the strongest and for rather large values of the exchange field and thicknesses of the AF layer the superconductivity can be completely suppressed. The amplitude of the oscillations is also weakly pronounced. It is due to the fact that the amplitude of the oscillating N\'eel triplets inside the AF layer is greatly suppressed in this case because of the overall suppression of superconductivity.

\begin{figure}[tb]
	\begin{center}
		\includegraphics[width=70mm]{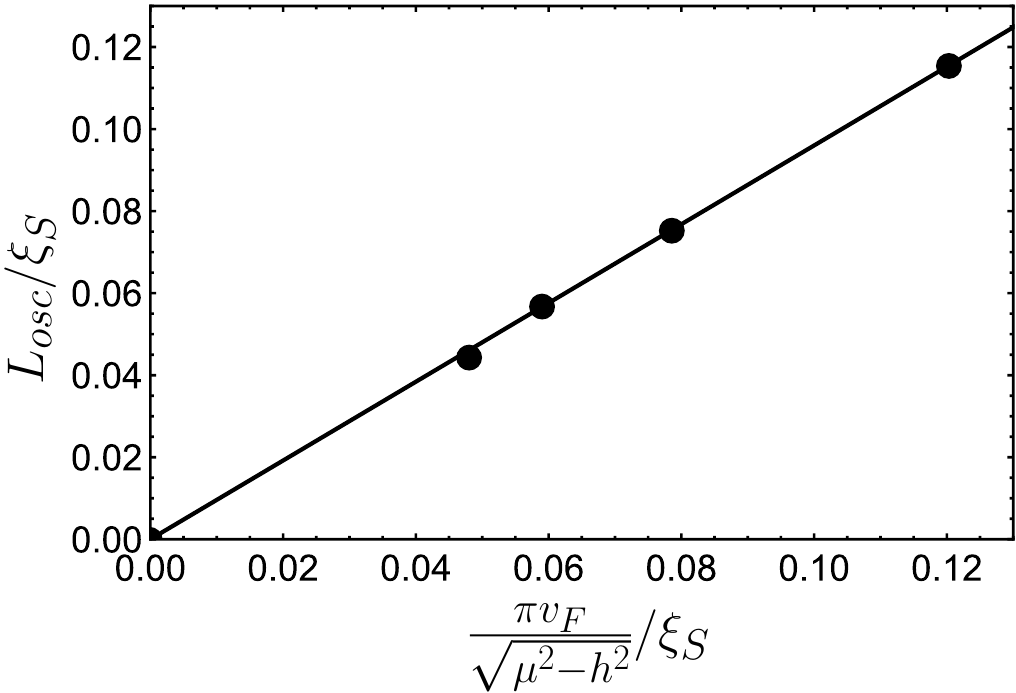}
		\caption{Period of oscillations $L_{osc}$ of the critical temperature for $h=0.2,0.6,0.8,0.92 \mu_{AF}$  as a function of $\frac{\pi v_F}{\sqrt{\mu_{AF}^2-h^2}}$.}
  \label{fig:period}
	\end{center}
\end{figure}
The period of the oscillations is described well by Eq.~(\ref{eq:period}) regardless of the thickness of the superconductor. The oscillation period extracted from the results presented in Fig.~\ref{fig:Tc2} is shown by black points in Fig.~\ref{fig:period}, which are in excellent agreement with Eq.~(\ref{eq:period}).

\section{Conclusions}
\label{conclusions}

In the present work an AF/S bilayer with metallic antiferromagnet is studied in the framework of the two-sublattice quasiclassical theory in terms of the Eilenberger equation for Green's functions. It is demonstrated that
the proximity-induced N\'eel triplet correlations decay into the depth of the antiferromagnet is superimposed by oscillations. The period of the oscillations is determined by the finite momentum of the N\'eel triplet pairs in the antiferromagnet caused by the Umklapp scattering at the AF/S interface. The oscillations manifest themselves in the oscillating dependence of the critical temperature on the AF thickness. 

The predicted oscillating behavior of the critical temperature is qualitatively similar to the oscillations observed in a number of experimental studies of the critical temperature of AF/S bilayers \cite{Bell2003,Hubener2002,Wu2013} and can provide a plausible explanation of these experimental findings. A quantitative comparison between the presented theory and experimental results is difficult because our model does not take into account influence of impurities and finite interface barrier. At the same time it is known that nonmagnetic impurities can provide additional suppression of superconductivity \cite{Fyhn2022_1} and, therefore, should also contribute to the experimental results. Nevertheless, our results demonstrate that the oscillating behavior of the critical temperature in AF/S bilayers can be a signature of the presence of proximity-induced N\'eel-type triplet correlations in the antiferromagnet.

\begin{acknowledgments}
The financial support from the Russian
Science Foundation via the RSF project No.22-22-00522 is acknowledged.    
\end{acknowledgments}

\section*{Appendix: Calculation of the normal state Green's function}

Let's define the following expansion of a Green's function in the basis of Pauli matrices in sublattice space:
\begin{align}
\check g_\sigma =\sum\limits_{i}\check g_{\sigma,i} \rho_i,~~~i\in \{0,x,y,z\}.
\end{align}
Then in the normal state, Eq. \ref{Eilenberger} takes the form:
\begin{align}
    &\pm i v \frac{d}{dx}\check g_{N,0}^\pm=0, \nonumber \\
    &2 h(x)\sigma \check g_{N,z}^\pm \pm i v\frac{d}{dx}\check g_{N,x}^\pm=0, \nonumber \\
    &2(\omega_m-i\mu(x))\check g_{N,z}^\pm \pm iv\frac{d}{dx}\check g_{N,y}^\pm=0, \nonumber \\
    &-2 \sigma h(x)\check g_{N,x}^\pm-2(\omega_m-i\mu(x))\check g_{N,y}^\pm \nonumber \\ 
    &\pm iv\frac{d}{dx}\check g_{N,z}^\pm=0 . 
    \label{eq:gN_components}
\end{align}
Here $h(x) = h\Theta(-x)$ and $\mu(x) = \mu_{AF}\Theta(-x) + \mu_S \Theta(x)$. Expanding normalization condition (\ref{eq:normalization_condition}) over the Pauli matrices we obtain:
\begin{align}
        \check g_{N,0}(x)=0,~\check g_{N,x}(x)^2+\check g_{N,y}(x)^2+\check g_{N,z}(x)^2=1.
\label{eq:nc_exp}
\end{align}
In the S region the solution of Eq.~(\ref{eq:gN_components}) accounting for the normalization condition takes the form:
\begin{align}
\begin{cases}
       \check g_{N,x}^S(x)=G \\
       \check g_{N,y}^S(x)=iDe^{\lambda x}-iEe^{-\lambda x} \\
       \check g_{N,z}^S(x)=De^{\lambda x}+Ee^{-\lambda x}
\end{cases}
\label{gS_app}
\end{align}
In AF region the solution takes the form:

\begin{align}
\begin{cases}
       \check g_{N,x}^{AF}(x)=&A+\frac{ih\sigma B}{\Lambda}e^{\beta x}+\frac{-ih\sigma C}{\Lambda}e^{-\beta x} \\
       \check g_{N,y}^{AF}(x)=&A\frac{h\sigma}{i(\mu_{AF}+i\omega_m)}+\frac{(\mu_{AF}+i\omega_m)B}{\Lambda}e^{\beta x}+ \\
       &\frac{-(\mu_{AF}+i\omega_m)C}{\Lambda}e^{-\beta x} \\
       \check g_{N,z}^{AF}(x)=&Be^{\beta x}+Ce^{-\beta x}
\end{cases},
\label{gAF_app}
\end{align}
where $\lambda=\frac{2(\omega_m-i\mu_{S})}{v}$, $\beta=\frac{2\Lambda}{v}$, $\Lambda=\sqrt{h^2-(\mu_{AF}+i\omega_m)^2}$, $A,B,C,D,E,G$ are unknown coefficients. Eq. \ref{eq:normalization_condition} takes the form:

\begin{align}
    4 BC+A^2\left (1-\frac{h^2}{(\mu_{AF}+i\omega_m)^2}\right)=1,~~~4DE+G^2=1.
\end{align}

To find the coefficients $A,B,C,D,E,G$, we should take into account two trajectories with $v_{F,x}=\pm v$. Boundary conditions at $x=+d_S$, $x=-d_{AF}$ and $x=0$ take the form:

\begin{align}
\begin{cases}
    \check g_N^{S,+}(x=+d_S)=\check g_N^{S,-}(x=+d_S) \\
    \check g_N^{AF,+}(x=-d_{AF})=\check g_N^{AF,-}(x=-d_{AF}) \\
    \check g_N^{S,\pm}(x=0)=\check g_N^{AF,\pm}(x=0)
\end{cases}
\label{bc_app}
\end{align}

Substituting Eqs.~(\ref{gS_app}) and (\ref{gAF_app}) into Eq.~(\ref{bc_app}) we obtain:

\begin{widetext}

\begin{align}
    B=\frac{h\sigma}{2\sqrt{B_0}} ~~ {\rm with}
\end{align}

\begin{align}
    B_0=(\mu_{AF}^2-h^2){\rm cosh}^2(\beta d_{AF})+i\mu_{AF}\sqrt{h^2-\mu_{AF}^2} 
    \times{\rm sinh}(2\beta d_{AF}){\rm coth}(\lambda d_S)+\nonumber \\(\mu_{AF}^2{\rm coth}^2(\lambda d_S)- 
    h^2 {\rm sinh}^{-2}(\lambda d_S))\sinh(\beta d_{AF}),
\end{align}

\begin{align}
    C=-B,~~~D=-B\frac{{\rm sinh}(\beta d_{AF})}{{\rm sinh}(\lambda d_S)},~~~E=-D,
\end{align}

\begin{align}
    A=\frac{B}{h\sigma}\left((-1+{\rm coth}(\lambda d_S))\left({\rm cosh}(\lambda d_S){\rm sinh}(\beta d_{AF})- \frac{i\mu_{AF}{\rm cosh}(\beta d_{AF}){\rm sinh}(\lambda d_S)}{\sqrt{h^2-\mu_{AF}^2}}\right)2e^{\lambda d_S}\mu_{AF}\right),
\end{align}

\begin{align}
    G=\frac{2B}{h\sigma}\left(\sqrt{h^2-\mu_{AF}^2}{\rm cosh}(\beta d_{AF})+\mu_{AF}{\rm coth}(\lambda d_{S}){\rm sinh}(\beta d_{AF})\right).
\end{align}

In case of $\mu_S=0$ (it is the case under consideration) solution for coefficients for anomalous Green's function $L^{\pm}_{1,\sigma},L^{\pm}_{2,\sigma}$ can be simplified and written here (the full solution can be also obtained, but it is rather long):

\begin{align}
L^{\pm}_{1,\sigma}-L^{\pm}_{2,\sigma}=2\frac{X {\rm cosh}(\hat\nu_1^\pm)-Y^\pm \frac{h \sigma\mu_{AF}}{\mu_{AF}^2-h^2}{\rm coth}^2(\hat\kappa_1^\pm){\rm sinh}(\hat\nu_1^\pm)-(Y^\pm h\sigma {\rm cosh}(\hat\nu_1^\pm)-X \mu_{AF}{\rm sinh}(\hat\nu_1^\pm))\frac{{\rm coth}(\hat\kappa_1^\pm)}{\sqrt{\mu_{AF}^2-h^2}}}{\left( {\rm cosh}(\hat\nu_1^\pm)+\frac{\mu_{AF}}{\sqrt{\mu_{AF}^2-h^2}}{\rm coth}(\hat\kappa_1^\pm){\rm sinh}(\hat\nu_1^\pm) \right)^2},
\end{align}
where 
\begin{eqnarray}
\hat\nu_1^{\pm}=\frac{\mp i \nu_1 d_S}{v}, ~~ 
\hat\kappa_1^{\pm}=\frac{\mp i \kappa_1 d_{AF}}{v}, ~~
X=-\frac{(\hat G_\sigma^\pm \hat s_1)}{\nu_1 |s_1|^2},~~
Y^{\pm}=-\frac{\mp i d_S}{v}~\frac{(\hat G_\sigma^\pm \hat s_4)}{|s_4|^2} .
\end{eqnarray}

\end{widetext}

\bibliography{oscillations}

\end{document}